\documentclass[prl,twocolumn,showpacs,amsmath,amssymb,superscriptaddress]{revtex4}

\usepackage{graphicx}
\usepackage{dcolumn}
\usepackage{bm}

\begin{document}

\title{Anomalous Resistance Ridges Along Filling Factor $\nu = 4i$}

\author{K. Takashina}
 \affiliation{NTT Basic Research Laboratories,
NTT Corporation, Atsugi-shi, Kanagawa 243-0198, Japan}

\author{M. Brun}
 \affiliation{NTT Basic Research Laboratories,
NTT Corporation, Atsugi-shi, Kanagawa 243-0198, Japan}

\author{T. Ota}
 \affiliation{ SORST JST, Kawaguchi, Saitama 331-0012, Japan}

\author{D.K. Maude}
 \affiliation{GHMFL, CNRS, BP 166, F-38042 Grenoble Cedex
9, France}

\author{A. Fujiwara}
 \affiliation{NTT Basic Research Laboratories,
NTT Corporation, Atsugi-shi, Kanagawa 243-0198, Japan}

\author{Y. Ono}
 \affiliation{NTT Basic Research Laboratories,
NTT Corporation, Atsugi-shi, Kanagawa 243-0198, Japan}

\author{Y. Takahashi}
\altaffiliation[Present Address: ]{Graduate School of Information
Science and Technology, Hokkaido University, Kita 14, Nishi 9,
Kita-ku, Sapporo, 060-0814 Japan}
 \affiliation{NTT Basic Research Laboratories,
NTT Corporation, Atsugi-shi, Kanagawa 243-0198, Japan}

\author{Y. Hirayama}
\altaffiliation[Present Address: ]{Graduate School of Science,
Tohoku University, 6-3 Aramakiaza Aoba, Aobaku, Sendai, 980-8578
Japan}
 \affiliation{NTT Basic Research Laboratories,
NTT Corporation, Atsugi-shi, Kanagawa 243-0198, Japan}
 \affiliation{ SORST JST, Kawaguchi, Saitama 331-0012, Japan}

\date{\today}

\begin{abstract}

We report anomalous structure in the magnetoresistance of
SiO$_2$/Si(100)/SiO$_2$ quantum wells. When Landau levels of
opposite valleys are driven through coincidence at the Fermi level,
the longitudinal resistance displays elevations at filling factors
that are integer multiples of 4 $(\nu=4i)$ accompanied by
suppression on either side of $\nu=4i$. This persists when either
magnetic field or valley splitting is swept leading to resistance
ridges running along $\nu=4i$. The range of field over which they
are observed points to the role of spin degeneracy, which is
directly confirmed by their disappearance under in-plane magnetic
field. The data suggest a new type of many-body effect due to the
combined degeneracy of valley and spin.

\end{abstract}

\maketitle


When two Landau Levels (LLs) of a two-dimensional electron system in
magnetic field ($B$)\cite{AndoFowlerSternReview} become degenerate
at the Fermi energy, many-body interactions come to the fore as the
electrons have relative freedom as to how they occupy the degenerate
levels. Recent research has revealed a variety of striking phenomena
that depend on the spin ($m_\mathrm{Z}$), orbital ($l$) and layer
(or confinement subband) indices of the LLs involved
\cite{Crossing,JunwirthMacDonald,ZeitlerPRL}.

In valley-degenerate silicon, further new physics can be expected
due to the addition of the valley degree of freedom, which differs
from the layer degree of freedom in that spatial wavefunctions of
different valleys are spatially coincident, rather like spatial
wavefunctions of states with differing spin. However, control over
the valley splitting ($\Delta_\mathrm{V}$) which lifts valley
degeneracy and also control over the spatial subbands in silicon
have only recently become readily accessible through the use of
SiO$_2$/Si/SiO$_2$ quantum well
structures\cite{ValleySIMOXOuisse,ValleySIMOX,ValleySIMOXZeroB}.
This is expected to offer new device possibilities and enable
otherwise inaccessible experiments to be performed, making further
characterization of these structures such as through
magneto-transport of urgent priority.

Here, we address the low-temperature magneto-transport properties of
SiO$_2$/Si/SiO$_2$ quantum wells when LLs are driven through
energetic coincidence by controlling $\Delta_\mathrm{V}$ and the
spatial subband separation. In transport measurements of other
systems, coincidences manifest themselves either by spikes or more
broad elevations in longitudinal resistance ($R_{xx}$) occurring
precisely when the LLs are degenerate and can shift as a function of
filling factor, although their visibility can change with underlying
total filling factor $\nu$. At $B$ lower than required for quantum
Hall physics, competing subbands or spin leads to beating patterns
or phase shifts in Shubnikov de Haas (SdH) oscillations reflecting
the density of states (DOS) at the Fermi level.

In this paper, under previously unexplored conditions attained in
SiO$_2$/Si/SiO$_2$ quantum wells, we observe periodic structure in
magneto-resistance where the rise and fall of $R_{xx}$ does not
correlate with any expected oscillations in the underlying density
of states. When LLs with differing $l$ and opposite valleys are
driven through coincidence, $R_{xx}$ is enhanced when $\nu$ is a
multiple of 4, accompanied by $R_{xx}$ suppression at the flanks.
This persists when either $B$ or $\Delta_\mathrm{V}$ is swept over
multiple orders of coincidences leading to resistance ridges running
along $\nu=4i$. The range of $B$ over which the ridges are observed
points to the role of spin degeneracy, which is directly
demonstrated by their disappearance under in-plane magnetic field.
The data suggest a new type of many-body effect due to the combined
degeneracy of valley and spin. The generality of this phenomenon and
the requirement of strong interactions is confirmed by measurements
at coincidences between LLs of different confinement subbands.


The LL coincidences are achieved by exploiting quantum wells made by
fabricating MOSFETs on two types of (100) Silicon-On-Insulator (SOI)
substrates [Fig.1(a,b)]. First is the ``SIMOX" (Separation by
IMplantation of OXygen) substrate where recent experiments have
shown that the buried oxide (BOX)-silicon interface allows access to
many orders of coincidence between LLs of opposite
valleys\cite{ValleySIMOXOuisse,ValleySIMOX,ValleySIMOXZeroB}. Second
is a bonded-SOI substrate, where both the BOX and the front-oxide
layers are formed through thermal oxidation\cite{UNIBOND}, leading
to quantum wells with relatively symmetric properties with respect
to electrostatic bias and $\Delta_\mathrm{V}$ in comparison to their
SIMOX counterparts. The mobility in both types of structure is
influenced by scattering from both interfaces which depends on the
potential bias. In a wide (25nm) SIMOX quantum well where the
electrons at the two interfaces are independent, electrons at the
front (back) interface have higher (lower) peak mobility of $1.0$
($0.4$)m$^2/$Vs\cite{ValleySIMOX}. Bonded-SOI samples had slightly
lower peak mobility at the front, while the back interface showed
values comparable to equivalent SIMOX samples.


\begin{figure}[t]
  \includegraphics[width=1\linewidth]{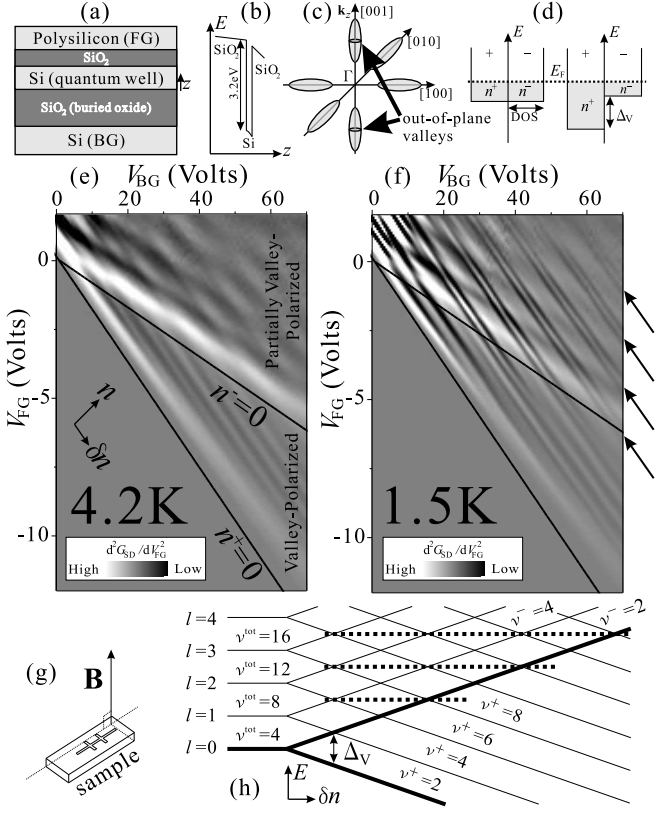}
  \caption{
    (a) Layer structure. (b) Quantum well potential. The electric field in the oxide layers are independently controlled by $V_\mathrm{FG}$ and $V_\mathrm{BG}$.
  (c) Only out-of plane valleys are occupied.
  (d) The resulting DOS, with and without valley splitting $\Delta_\mathrm{V}$.
  (e) Gray-scale plot of d$^2G_{\mathrm{SD}}/$d$V^2_{\mathrm{SD}}$ at $B=5.5$T,
  $T=4.2$K. Solid lines labeled $n^+=0$ and $n^-=0$ mark lines of zero electron density in the
  lower and upper valley subbands as extracted using $\alpha=0.46$meV$/10^{15}$m$^{-2}$.
  (f) Data at 1.5K. Ridges along $\nu=16$, 20, 24 and 28 are marked by
  arrows. Solid lines are the same as those in (e). (g) Sample
  configuration.
  (h) Diagram of the spin degenerate LLs as a function of potential
  bias $\delta n$. ($\Delta_\mathrm{V}\propto \delta n$).
  Thicker lines represent the lowest LLs for the two valley subbands + and -.
  Dotted lines schematically mark where the ridges are observed.}
\end{figure}

We first focus on the phenomenology in a narrow SIMOX quantum well
with a nominal thickness of 8nm, where only the lowest confinement
subband of the out-of-plane valleys [Fig.1(c)] is
occupied\cite{ValleySIMOXZeroB,IEEESubbands}. Transport data at
$B=5.5$T, 4.2K are shown in Fig. 1(e). Since the concentration and
mobility change a great deal with front- and back-gate voltages
($V_{\mathrm{FG}}$ and $V_{\mathrm{BG}}$) which leads to strong
variations in resistivity, we plot the double differential of the
conductance d$^2G_{\mathrm{SD}}/$d$V^2_{\mathrm{SD}}$ in order to
highlight SdH oscillations over a wide range of
$(V_{\mathrm{FG}},V_{\mathrm{BG}})$.

The total electron concentration $n$ is given by
$n=C_\mathrm{F}(V_\mathrm{FG}-V^\mathrm{Th}_\mathrm{F})+C_\mathrm{B}(V_\mathrm{BG}-V^\mathrm{Th}_\mathrm{B})$
where $C_\mathrm{F}=485\mu$Fm$^{-2}$ and
$C_\mathrm{B}=92\mu$Fm$^{-2}$ are front- and back-gate capacitances,
and $V^\mathrm{Th}_\mathrm{F}$ and $V^\mathrm{Th}_\mathrm{B}$ are
constant offsets. The occupation of the upper and lower valley
subbands [Fig.1(d)] are approximated by a phenomenological formula
for the valley splitting: $\Delta_\mathrm{V} = \alpha \delta n$ when
$\delta n>0$; $\delta n = n_{\mathrm{B}}-n_{\mathrm{F}}$,
$n_{\mathrm{B}}$ and $n_{\mathrm{F}}$ being electron densities
contributed by the back and front gates, respectively. The valley
factor $\alpha$ for the Si-buried oxide interface in this sample has
a large value of about $0.46$meV$/10^{15}$m$^{-2}$
\cite{ValleySIMOXZeroB}. Corresponding LLs are schematically
depicted in Fig. 1(h).

At 4.2K, in the completely valley-polarized region where only the
lower valley is occupied at large positive $V_{\mathrm{BG}}$ and
large negative $V_{\mathrm{FG}}$ [between the two solid lines in
Fig. 1(e)], oscillations occur with a period of $\Delta \nu^+=2$ due
to two-fold spin degeneracy since the spin splitting is not
resolved. In the partially valley-polarized region, two sets of
oscillations can be seen, where one set corresponds to electrons in
the upper valley subband with a corresponding period of $\Delta
\nu^-=2$.


When the temperature is reduced to 1.5K, oscillations remain
qualitatively unchanged where only one valley subband is occupied
and the upper valley subband is far above the Fermi energy [bottom
right of Fig 1(f)]. Surprisingly, however, when the upper valley
subband becomes occupied, the pattern is drastically altered to one
where by far the strongest features (some marked by arrows) run
parallel to lines of constant total filling factor
$\nu^{\mathrm{tot}}=\nu^++\nu^-$. The transport no longer reflects
the expected oscillations in the single particle DOS [Fig. 1(h)].


\begin{figure}[b]
  \includegraphics[width=0.95\linewidth]{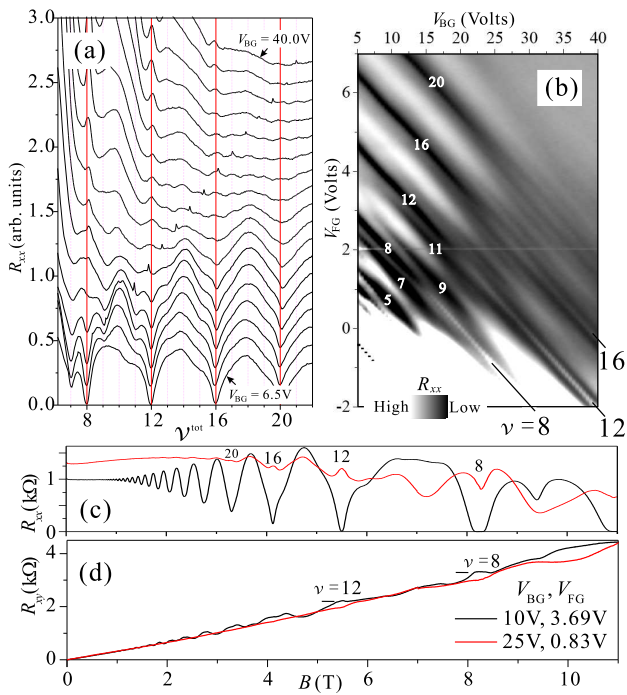}
  \caption{ (a) $R_{xx}$ as a function of filling
  factor $\nu$ at various values of $V_{\mathrm{BG}}$ (fixed intervals, data offset)
  at $B=5.5$T, 350mK.
  (b)Gray-scale plot of the same data as (a). A monotonic background
  has been subtracted for clarity. White numbers indicate filling factors.
  (c) and (d)$R_{xx}$ and $R_{xy}$($B$) taken at different $V_\mathrm{BG}$ keeping $n$ constant.
  Ridges are seen at $\nu=12$, 16 and 20 but not 8 for $V_\mathrm{BG}=$25V, $V_\mathrm{FG}=$0.83V.}
\end{figure}

Since the conductance $G$ does not necessarily reflect the
resistivity of the 2DES, we also present results of four-terminal
measurements. Figures 2(a) and (b) show $R_{xx}$ measured at a lower
temperature of 350mK where there is less thermal broadening. The
plot shows data up to a larger value of $V_\mathrm{FG}$ (compared to
Fig.1 (e,f)) to examine the behavior when $\Delta_\mathrm{V}$ is
increased from its usual small value. When $\Delta_\mathrm{V}$ is
much smaller than $\hbar \omega_\mathrm{C}$ at small values of
$V_{\mathrm{BG}}$, deep minima are seen when $\nu$ is a multiple of
four. As $\Delta_\mathrm{V}$ is increased, adjacent LLs of opposite
valleys begin to overlap and these minima are replaced by peaks.
Minima at odd $\nu$ are also seen corresponding to valley gaps
\cite{ValleySIMOX}. With $V_{\mathrm{BG}}$ increased further,
features that can be attributed to individual energy gaps between
LLs disappear and the only remaining sharp features occur near
$\nu^\mathrm{{tot}}=4i$. Specifically, there are sharp elevations in
the resistance when $\nu^\mathrm{tot}=4i$, accompanied by resistance
suppression slightly off $\nu^\mathrm{tot}=4i$. This behavior
persists when $\Delta_\mathrm{V}$ is swept beyond a particular
coincidence to the adjacent order coincidence, leading to resistance
ridges running along $\nu^\mathrm{tot}=4i$.


The ridge structure is evidently a consequence of overlap between
LLs of opposite valleys since more usual SdH oscillations are
recovered at complete valley polarization [Fig. 1(f)]. However, it
cannot be explained by a simple picture of overlapping LLs depicted
in Fig. 1(h) with $R_{xx}$ reflecting the single particle DOS. Nor
is it simply enhanced or accentuated at specific LL
coincidences\cite{Crossing}. Furthermore, it can neither be
explained by simple considerations of anti-crossing. However, the
result that the structure depends strongly and consistently on the
total electron concentration despite major rearrangements of the
underlying single particle energies suggests a many-body origin due
to electron-electron interactions, where the energies involved are
larger than or comparable to the separation between single-particle
LLs\cite{Ridges under n-=0}.

One possible clue as to the origin is the absence of much structure
between the ridges despite the low temperature. It suggests that $B$
is low enough and the disorder great enough for up and down spin
states to be effectively degenerate. We have found that at higher
$B$, these resistance ridges disappear, accompanied by the
appearance of structure attributable to individual spin-resolved LLs
\cite{disappearanceWithField}. This leads us to suspect that spin
degeneracy plays an important role in the formation of the ridges.


\begin{figure}[b]
  \includegraphics[width=1\linewidth]{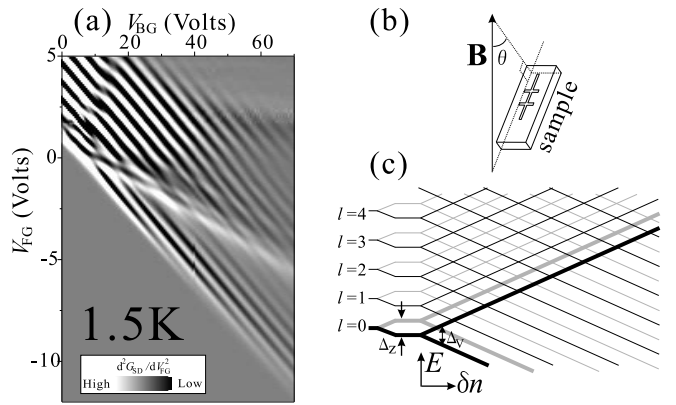}
  \caption{(a) d$^2G_{\mathrm{SD}}/$d$V^2_{\mathrm{SD}}$ at 1.5K, $\theta=55.6$ degrees,
  $\mathbf{B}=9.735$T, $B_{\bot}=5.5$T. (b) Direction of the sample with respect to field. (c) Schematic diagram of spin-split LLs.}
\end{figure}

Evidence for the requirement of spin degeneracy for observing the
ridges comes from in-plane field measurements. Figure 3(a) shows
d$^2G_{\mathrm{SD}}/$d$V^2_{\mathrm{SD}}$ at 1.5K, when the sample
is tilted by an angle of $\theta=55.6$ degrees [Fig.3(b)] with a
total field of 9.735T keeping the component perpendicular to the
sample at 5.5T (the value for Figs. 1, 2(a,b)). The Zeeman energy
($\Delta_\mathrm{Z}$) is increased (from 0.6 to 1.1 meV taking a
$g$-factor of 2) while maintaining $\hbar\omega_{\mathrm{C}}$ (at
3.3 meV taking $m^*=0.19 m_0$). $\Delta_\mathrm{V}$ is unaffected by
in-plane field \cite{InPlaneFieldonValleys}. It is clearly seen that
the anomalous ridges completely disappear with large in-plane field,
being replaced by periodic oscillations. Although the oscillations
do not simply follow the single particle picture depicted in Fig.
3(c), presumably due to residual interactions, there is no structure
indicating special behavior at $\nu^\mathrm{tot}=4i$, demonstrating
that large Zeeman energy prevents the formation of the anomalous
ridges. In turn, this suggests that the ridges occur due to a
combined degeneracy of valley and spin. i.e. at coincidences between
spin degenerate LLs of opposite valleys.


\begin{figure}[t]
  \includegraphics[width=1\linewidth]{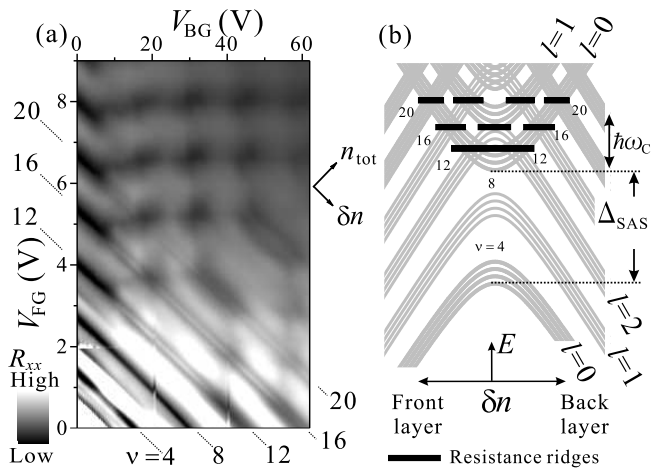}
  \caption{(a) $R_{xx}$
of the bonded-SOI sample (8T, 350mK). (b) Evolution of the lowest
six orbital LLs from the front layer to those of the back.
($\Delta_{\mathrm{SAS}}$ has been adjusted for each orbital to
reflect its dependence on concentration.)}
\end{figure}

The generality of the ridges and the requirement of
electron-electron interactions is demonstrated by measurements of
coincidences between levels originating from different confinement
subbands. These are examined on the second type of structure
fabricated on bonded-SOI substrates whose properties are relatively
symmetric with respect to potential bias. The well is nominally 11nm
thick, bringing the second confinement subband into
play\cite{ValleySIMOX,IEEESubbands}. When $V_{\mathrm{BG}}$
($V_{\mathrm{FG}}$) is zero, there are no electrons at the back
(front) Si/SiO$_{2}$ interface leading to a single set of SdH
oscillations[Fig.4(a)]. At $\nu = 4$ and 8, the SdH minima are
continuous between the regions explicitly associated with the front
and back due to the energy separation between the lowest and the
first excited subband ($\Delta_{\mathrm{SAS}}$) being of the order
of $2\hbar\omega_\mathrm{C}$. At large values of $V_{\mathrm{FG}}$,
and $V_{\mathrm{BG}}$ ($V_{\mathrm{FG}}> 6$V,
$V_{\mathrm{BG}}>10$V), a square lattice pattern is seen
corresponding to independent SdH oscillations. When the filling
factors at the front and back are the same or similar ($\delta n
\sim 0$), some minima are absent and some extended, similar to GaAs
double-quantum wells\cite{MurakiSoldStatCommun} except that each LL
is 4-fold degenerate.

In between the high $\nu$ bi-layer behavior and the low $\nu$
single-layer behavior, the data reveal clear resistance ridges
running along total filling factor $\nu = 4i$ (12, 16 and 20) with
minima along their sides, in a very similar manner as for the SIMOX
sample [Fig.2(b)]. The ridges occur along specific values of total
filling factor as opposed to the filling factor of the front or back
layer showing that this is an effect arising from the collaboration
of carriers from both layers.

The schematic diagram [Fig.4(b)] shows the qualitative behavior of
spin and valley degenerate LLs. LLs of the two confinement subbands
contain levels from opposite valleys which implies that their
crossing involves coincidence between LLs of opposite valleys. These
ridges are therefore likely to be consequences of the same mechanism
as those observed at coincidences between LLs of opposite valleys
but from the same confinement subband in the SIMOX sample. This is
also corroborated by the result that the ridges occur at $\nu=4i$
despite the additional multiplicity due to the second confinement
subband. The range of magnetic field over which the ridges are
observed is also similar to the SIMOX sample. At low enough magnetic
field the ridges become unresolved, and at high field, as individual
spin-split LLs resolve, they compete with the formation of quantum
Hall states which suppresses resistance. Within this field-range,
the ridges appear whenever there are coincidences or overlap between
bonding and antibonding levels at $\nu = 4i$. i.e., In Fig. 4(a),
there are no ridges along $\nu=4$ and 8 since only states of the
bonding (ground) subband are occupied and there are no level
crossings. Along filling factors $\nu=12$, 16 and 20, they appear
where there are level crossings and disappear where the energy gaps
are large, forming quantum Hall states with suppressed $R_{xx}$.

The dependence of the ridges on electron concentration may also be
instructive. Ridges at $\nu=12$ and $\nu=16$ appear strongly but the
ridge along $\nu=20$ is weaker and at higher filling factors the
ridges are no longer visible. This concentration dependence is
likely to be related to the overlap in real-space of the coincident
states. Increased concentration increases the distance between the
front and back-layers due to band-bending, also reflected by smaller
$\Delta_{\mathrm{SAS}}$, leading to weaker interactions between
their electrons. Indeed, our measurements on samples with nominal
quantum well thicknesses greater than 14nm showed no ridges at
coincidences between confinement subbands also demonstrating the
requirement of strong interactions.


In conclusion, we have shown anomalous ridge structures to appear in
resistivity at certain conditions of LL coincidences. In-plane field
measurements show the vital role of spin degeneracy while
coincidences between levels of confinement subbands demonstrate the
role of interactions and the generality of the phenomenon. Our data
point to the combined degeneracy of valley and spin as the cause,
suggesting a new type of many body effect at coincidences between
more than two LLs.

We are deeply grateful for helpful discussions with Norio Kumada and
Koji Muraki. This work is partially supported by JSPS KAKENHI
(16206003) and (16206038).


\begin{thebibliography}{0}
\bibitem{Ridges under n-=0} The persistence of the ridge strucure
below the line $n^-=0$ may be reflecting the energy scale of the
interactions.
\bibitem{disappearanceWithField} The
ridges disappear at lower field for lower asymmetry of the quantum
well. We expect that this is also related to the change in disorder
with electric field.
\end{thebibliography}
\end{document}